\documentclass[reqno,12pt]{article}
\usepackage{amsmath} 
\usepackage{bm}
\usepackage{a4wide,amssymb}
\usepackage{graphicx,xcolor}
\usepackage{epstopdf}
\usepackage{bbm}

\renewcommand{\(}{\left(}          \renewcommand{\)}{\right)}
\renewcommand{\[}{\left[}          \renewcommand{\]}{\right]}
 \newcommand{\nn}{\nonumber}
\newcommand{\R}{{\mathbb R}}

\newcommand\om{\omega}

\newcommand{\e}{\varepsilon}

\newcommand{\pa}{\partial}

\newcommand{\la}{\lambda}

\def\G{\Gamma}

\def\p{{\pi}}

\def\bz{{\bf z}}

\def\bv{{\bf v}}

\def\RRR{{\mathbb R}}

\def\CC{{\cal C}}
\def\SS{{\cal S}}

\newtheorem{thm}{Theorem}

\def\be{\begin{equation}}
\def\ee{\end{equation}}
\def\bea{\begin{eqnarray}}
\def\eea{\end{eqnarray}}
\def\ni{\noindent}
\def\nn{\nonumber}

\def\ol{\overline}

\def\o{\omega}
\def\b{\beta}

\begin{document}


\begin{center} 
{\bf \Large{Backward Clusters, Hierarchy and Wild Sums for a Hard Sphere System in a Low--Density Regime}}

\vspace{1cm}
{\large K. Aoki$^{1}$, M. Pulvirenti$^{2}$, S. Simonella$^{3}$, T. Tsuji$^{1}$}

\vspace{0.5cm}
{$1.$\scshape { Department of Mechanical Engineering and Science, Kyoto University\\
Katsura, 615-8540 Kyoto -- Japan \\ \smallskip
$2.$\small \ Dipartimento di Matematica, Universit\`a di Roma La Sapienza\\ 
Piazzale Aldo Moro 5, 00185 Roma -- Italy \\ \smallskip
$3.$\ Zentrum Mathematik, TU M\"{u}nchen \\ Boltzmannstrasse 3, 85748 Garching -- Germany
}}
\end{center}

\vspace{0.5cm}
ABSTRACT. We study the statistics of backward clusters in a gas of hard spheres
at low density. A backward cluster is defined as the group of particles involved directly
or indirectly in the backwards--in--time dynamics of a given tagged sphere. We derive upper and lower bounds
on the average size of clusters by using the theory of the homogeneous Boltzmann equation
combined with suitable hierarchical expansions. These representations are known in the easier context of 
Maxwellian molecules (Wild sums). We test our results with a numerical experiment based on molecular 
dynamics simulations.

\vspace{0.3cm}

{\em }

\thispagestyle{empty}


\section{Introduction} \label{sec:intro}
\setcounter{equation}{0}    
\def\theequation{1.\arabic{equation}}

Consider a system of $N$ 
identical hard spheres of diameter $\e$ moving in the whole space $\RRR^3$
or in a bounded box with reflecting boundary conditions. The collisions between spheres are
governed by the usual laws 
of elastic reflection. We order the particles with an index $i=1,2,\cdots,N$.
A configuration of the system is $\bz_N=( z_1, \cdots, z_N )$, where $z_i=(x_i,v_i)$ are the position 
and the velocity of particle $i$  respectively. Let us assign a probability density $W_0^N$ on the $N-$particle phase space,
assuming it symmetric in the exchange of the particles, and let $W^N(t)$ be its time 
evolution according to the hard sphere dynamics.
Finally, for $j=1,2,\cdots,N$, denote by $f^N_{0,j}$ and $f^N_{j} (t) $ the 
$j-$particle marginals of $ W_0^N$ and $W^N(t)$ respectively.

Given a tagged particle, say particle $1$, consider $z_1 (t, \bz_N)$ its state 
(position and velocity) at time $t$  for the initial configuration $\bz_N$.
We define the {\em backward cluster} of particle $1$ (at time $t$ and for the initial configuration $\bz_N$) as
the set of particles with indices  $J \subset I_N$, where $I_N=\{ 1,2, \cdots , N\}$, constructed in the following way.
Going back in time starting from $z_1 (t, \bz_N)$, let $i_1$ be the (index of the) first particle colliding with $1$. Next, considering the two particles $1$ and $i_1$,  let us go back in time up to the first collision of one particle of the pair with a new particle $i_2$ and so on up to time $0$.
Then  $J=\{ i_1,i_2, \cdots , i_n\}$ with $i_r \neq i_s$ for $r \neq s$.
We denote by $K$ the cardinality of $J$, i.e., $K=|J|$.

In this paper we are interested in studying  the quantity $\langle K \rangle_t$ that is the average (with respect to the initial distribution)
of the cardinality of the backward cluster of a tagged particle at time $t$.  In a general context this is a hard task,
however we limit ourselves in considering  $\langle K \rangle_t$ in a low--density situation, namely in the 
{\em Boltzmann--Grad limit} \cite{Gr49,Gr58}
\be
N  \to \infty, \quad \e \to 0 \quad \mbox{and} \quad N\e^2 \to \la^{-1} >0\;,
\ee
where $\la$ is a constant proportional to the mean free path. We fix $\lambda = 1$ in the rest of the paper. 

Moreover, we shall assume that the initial distribution is approximately factorized, namely the marginals of 
the initial distributions do factorize in the Boltzmann--Grad limit, and that the one particle distribution is 
independent of $x$. 

In this situation 
it is believed (and in fact proved for short times and under suitable uniform estimates on the $f^N_{0,j}$)
that the system is ruled by the homogeneous Boltzmann equation, which we remind here for the unknown 
$f=f(v,t)$:
\be
\pa_t f(v,t)
= \int_{\RRR^3\times S^2_+} dv_1 d\o \ (v-v_1)\cdot\o \Big\{f(v_1',t)f(v',t)-
f(v_1,t)f(v,t)\Big\}
\label{BE}
\ee
where $S_+^2=\{\om \in S^2 |\ (v-v_1)\cdot\o \geq 0\},$ $S^2$ is the unit sphere in $\RRR^3$ 
(with surface measure $d\o$), $(v,v_1)$ is a pair of 
velocities in incoming collision configuration   and $(v',v_1')$ is the corresponding pair of outgoing 
velocities defined by the elastic reflection rules
\be
\begin{cases}
\displaystyle v'=v-\om [\om\cdot(v-v_1)] \\
\displaystyle  v_1'=v_1+\om[\om\cdot(v-v_1)]
\end{cases}\;.\label{eq:coll}
\ee
The Cauchy theory of equation \eqref{BE} is well known, see \cite{V02} and
references quoted therein. The solution $f = f(v,t)$
is usually interpreted as the one--particle distribution of the system in the low--density regime.

In the next section we will give a precise definition for the quantity $\langle K \rangle_t$ associated
to the Boltzmann equation \eqref{BE}. In order to describe the long time behaviour of $\langle K \rangle_t$, 
it will be convenient to focus on:
\be
r=\lim_{t \to \infty} \frac 1 t \log \langle K \rangle_t\;.
\label{rate}
\ee
In Section \ref{sec:Max} we shall compute exactly $\langle K \rangle_t$ and $r$ for a simplified model of 
Maxwellian molecules. In this case $r$ is an absolute constant. 
In Sect. \ref{sec:estimate} we come back to the hard sphere system and prove an exponential estimate of 
the growth in time of $\langle K \rangle_t $. However the rate is not constant anymore, but depends on the 
initial datum. 

A comparison of $\langle K \rangle_t $, $r$ with the corresponding quantities at the level of the particle system 
($\e>0$) will be performed numerically in Section \ref{sec:Numsim}.

\bigskip
The cluster dynamics outside the low--density regime
has been studied previously, both analytically \cite{S73,S74} 
as regards the equilibrium dynamics of infinite particle systems,
and numerically \cite{GKBSZ08}. In the latter reference, 
recent applications to several domains are mentioned and 
discussed, such as plasma physics, geophysics or economics.

We stress however that the notion of cluster introduced in 
these papers (see e.g. \cite{GKBSZ08}, Section 2) differs from 
that of ``backward cluster'' considered in the present work . 
This refers exclusively to the backward dynamics of one single 
tagged particle. In particular, note that the particles join a backward cluster 
{\em one by one}. In other words, when particle $i$ joins the backward cluster of 
particle $1$, the particles belonging to the backward cluster of $i$, not involved in the backward cluster of $1$, are ignored. 

This concept emerges naturally from the perturbative description of the
Boltzmann--Grad limit, as enlightened in the following section,
and is related to the Markovian character of the dynamics.

\bigskip
We conclude this introduction by observing that the interest on the control of the backward cluster is also 
related to the problem of ``mathematical validity of the Boltzmann equation''. 
It is known that the validity of the Boltzmann equation is crucially dependent on the factorization of the marginals 
of the $N-$particle system $f^N_j(\bv_j,t)$, where $\bv_j= (v_1, \cdots, v_j )$, at any positive time $t$. In order 
that this property is fulfilled, it is necessary that 
the backward clusters of any 
couple of particles (say $1$ and $2$) are disjoint. When such two clusters are finite, the probability that the two particles 
are dynamically correlated is $O\left( \frac {\langle K \rangle_t^2} {N}\right)$.  We estimate in Sect. \ref{sec:estimate} 
$\langle K \rangle_t$ assuming that the Boltzmann--Grad limit has been achieved. Therefore this result can be 
interpreted as a compatibility argument.

Another connected problem is the following. Even though the convergence $f^N_j (\bv_j,t) \to f(v_1,t)f(v_2,t)\cdots f(v_j,t)$ 
in the Boltzmann--Grad limit has been proven at least for short times \cite{La75}, one can wonder for which $j$  
the asymptotic equivalence holds.
If the $j$ particles have  finite backward clusters, we can argue that the probability of correlations between any pair 
in the group of $j$ particles is $O(\frac {j^2} N)$. Therefore we expect that the factorization property of marginals holds 
when $\lim_N \frac {j^2}{N}=0$. Actually in \cite{PS14} it has been proven that the propagation of chaos holds
for short times if $j\leq N^{\alpha}$ for $\alpha$ small enough.
 
Finally it may be worth noting that the notion of backward cluster could be of interest in problems of population dynamics 
where one is interested in the mean growth of a group of individuals which contacted, directly or indirectly, a given one.

\section{Preliminaries: the Boltzmann-Grad limit} \label{sec:pre}
\setcounter{equation}{0}    
\def\theequation{2.\arabic{equation}}

In what follows we expand the solution of \eqref{BE}, i.e. $f(v,t)$, in terms of a sum 
\be
f=\sum_{n=0}^\infty f^{(n)}\;,
\label{exp}
\ee
where $f^{(n)}$ is interpreted as the contribution to the probability density $f$ due to the event: 
{\it the backward cluster of $1$ has cardinality $n$}. 

Let $f_0 = f_0(v)$ be the initial datum for the Boltzmann equation.
By \eqref {BE} it follows naturally that
\be
f^{(0)}(v,t) = e^{-\int_0^t ds R (v,s) } f_0(v),
\label{0}
\ee
where
\be
R (v,t)=  \int_{\RRR^3\times S^2_+} dv_1 d\o \ (v-v_1)\cdot \o f(v_1,t)=\pi  \int_{\RRR^3} dv_1   |v-v_1| f(v_1,t)\;.
\label{nu}
\ee

Before giving the other terms of the expansion we introduce a useful tool, namely the Boltzmann hierarchy.

Suppose that $f$ is a solution to the Boltzmann equation \eqref{BE} and consider the products
\be
f_j(\bv_j,t)=f(t)^{\otimes j}(\bv_j) = f(v_1,t)f(v_2,t)\cdots f(v_j,t)\;  \label{eq:fjtdef},
\ee
where $\bv_j= (v_1, \cdots, v_j )$.
The family of $f_j$  solves then the hierarchy of equations
\be
 \pa_t f_j (\bv_j,t) = \CC_{j+1}f_{j+1}(\bv_j,t)-R_j (\bv_j,t) f_j(\bv_j,t) \;,
 \label{hie}
\ee
where
\bea
&& \CC_{j+1} =\sum_{k=1}^j \CC_{k,j+1} \label{eq:defBco} \\
&& \CC_{k,j+1}f_{j+1}(\bv_j,t) = \int_{\RRR^3\times S^2_+}dv_{j+1}d\o\, 
(v_k-v_{j+1})\cdot\o \, f_{j+1}(v_1,\cdots,v'_k,\cdots,v_j ,v'_{j+1},t)\;, \nn
\eea
\bea
&& \begin{cases}
\displaystyle v'_k=v_{k}-\om [\om\cdot(v_k-v_{j+1})] \\
\displaystyle  v'_{j+1}=v_{j+1}+\om[\om\cdot(v_k-v_{j+1})]
\end{cases}\;, \\
&& \nn\\
&& S^2_+ = \{\o\ |\ (v_k - v_{j+1})\cdot\o \geq 0\}\;,\nn 
\eea
and
\be
R_j (\bv_j,t) = \sum_{k=1}^j R(v_k,t)\;.
\ee

By using a formal solution of \eqref {hie} iteratively, we can express $f_j(t)$ via the following series
\bea
&& f_j(t)= \sum_{n\geq 0}\int_0^t dt_1 \int_0^{t_1} dt_2 \cdots \int_0^{t_{n-1}}dt_n \nn\\
&& \ \ \ \ \ \ \ \ \ \ \ \ \ \ \ \ \cdot\SS_j(t_1,t)\CC _{j+1}\SS_{j+1}(t_2,t_1)\cdots \CC _{j+n}\SS_{j+n}(0,t_n) f_{0}^{\otimes (j+n)}\;,
\label{eq:fjexp}
\eea
where we use the conventions $t_0 = t, t_{n+1}=0$, and the term $n=0$ should be interpreted as 
$\SS_{j}(0,t) f_{0}^{\otimes j}$. Here $\SS_j(t_1,t_2)   $ is the  multiplicative operator defined as
\be
\SS_j(t_1,t_2)f_j(\bv_j,\cdot) = e^{-\int_{t_1}^{t_2} ds R_j (\bv_j,s) }f_j(\bv_j,\cdot)\;. \label{eq:ffodef}
\ee
Note that in \eqref{eq:fjexp}, and in the formulas below, the dependence on $\bv_j$ is not shown explicitly.

In more detail,
\bea
&& f_j(t)= \sum_{n\geq 0}\sum_{k_1, \cdots, k_n} \int_0^t dt_1 \int_0^{t_1} dt_2 \cdots \int_0^{t_{n-1}}dt_n \nn\\
&& \ \ \ \ \ \ \ \ \ \ \ \ \ \ \ \ \cdot\SS_j(t_1,t)\CC _{k_1,j+1}\SS_{j+1}(t_2,t_1)\cdots \CC _{k_n, j+n}\SS_{j+n}(0,t_n) f_{0}^{\otimes (j+n)}\;,
\label{eq:fjexp1}
\eea
where $k_1\in \{1, \cdots, j \} , k_2\in \{1,\cdots, j+1\}, \cdots, k_n \in \{1,2, \cdots, j+n-1\}$. 
We call any sequence $\{ k_1, \cdots, k_n\}$ of this type a ``$j-$particle tree with $n$ creations''. 
Indeed any new created particle in formula \eqref{eq:fjexp1}, say $j+r$,  can be attached to any 
of the previous $j+r-1$ particles (for more details on this representation, see e.g. \cite{PS14}). 
We denote a $j-$particle tree with $n$ creations by $\G_n(j)$.

Fixed $\Gamma_n(j) $, $\o_1, \cdots, \o_n$, $\bv_j$ and the velocities of 
the new particles $v_{j+1}, \cdots, v_{j+n}$, we introduce a sequence of vector velocities 
$\bv^s$, $s=0, \cdots, n$, by setting:
$$
\bv^0=\bv_j,    \quad \bv^s=( v_1^{s-1},\cdots, v_{k_s}', \cdots,v_{j+s-1}^{s-1}, v_{j+s}' )   \quad s\geq 1
$$
where, at step $s$,  the pair  $v_{k_s}'  , v_{j+s}'$ are the pre--collisional velocities 
(in the collision with impact vector $\o_s$) of the pair $v_{k_s}^{s-1},  v_{j+s}^{s-1}$ 
(which are, by construction, post--collisional). This allows to write
\eqref{eq:fjexp1} more explicitly as
\bea
\label{eq:fjexp2}
&& f_j(t)= \sum_{n\geq 0}\sum_{\Gamma_n (j)}  \int_0^t dt_1  \cdots \int_0^{t_{n-1}}dt_n \int_{\RRR^{3n}} 
dv_{j+1} \cdots dv_{j+n} 
e^{-\int_{t_{1}}^{t} ds\, R_{j}(\bv_{j},s)}\,
 \nn \\ 
&& \ \ \ \left(\prod_{r=1}^n  
\int_{(v^{r-1}_{k_r} -v_{j+r})\cdot\o_r  \geq 0} d\o_r \, (v^{r-1}_{k_r} -v_{j+r})\cdot\o_r \,
e^{-\int_{t_{r+1}}^{t_{r}} ds\, R_{j+r}(\bv^{r},s)}\right)
f_{0}^{\otimes(j+n)} (\bv^{n})\;, \nn\\
\eea
where we are using the convention $t_{n+1}=0$.

Formula \eqref{eq:fjexp} expresses the solution to the Boltzmann equation \eqref{BE} 
in terms of an expansion on the number of collisions. Each term of the series is the contribution 
to $f_j$ due to the event in which the first $j$ particles and the collided particles in the backward 
dynamics deliver exactly $n$ collisions. Setting $f_j=\sum_n f^{(n)}_j$, we identify
\bea
&& f_j^{(n)}(t) =  \int_0^t dt_1  \cdots \int_0^{t_{n-1}}dt_n  \, \label{jn}\\
&& \cdot e^{-\int_{t_1} ^t ds R_j (s) } \CC_{j+1}   e^{-\int_{t_2} ^{t_1} ds R_{j+1}(s) } \CC_{j+2}
\cdots  \CC_{j+n} \, e^{-\int_0^{t_n} 
ds R_{j+n}(s) } f_0^{\otimes(j+n)}\;.\nn
\eea
In particular \eqref{exp} holds with 
\bea
&& f^{(n)}(t) = \int_0^t dt_1  \cdots \int_0^{t_{n-1}}dt_n \label{n} \\
&& \cdot e^{-\int_{t_1} ^t ds 
R_1 (s) } \CC_{2}   e^{-\int_{t_2} ^{t_1} ds R_{2}(s) } \CC_3
\cdots \CC_{1+n}e^{-\int_0^{t_n} ds R_{1+n}(s) } f_0^{\otimes(1+n)}\;.\nn
\eea
As a consequence, we define
\be
\langle K \rangle_t=\sum_{n=0}^{\infty}  n \int dv \, f^{(n)} (v,t).
\label{meancl}
\ee

A remarkable  property which will be used later on is the following:
\be
f_2^{(n)}=\sum_{\substack{n_1,n_2 :\\ n_1+n_2=n}} f^{(n_1)} f^{(n_2)}\;.
\label{prod}
\ee
This is consequence of the rather obvious identity
\be
\label {prod1}
\sum_{\Gamma_n (2)}  \int_0^t dt_1  \cdots \int_0^{t_{n-1}}dt_n=\sum_{\substack {n_1, n_2: \\ n_1+n_2=n}} \,
\sum_{\substack{\Gamma_{n_1}(1)\\ \Gamma_{n_2} (1)}} \, \int_0^t dt_1^1  \cdots \int_0^{t^1_{n_1-1}}dt^1_{n_1}
 \int_0^t dt^2_1  \cdots \int_0^{t^2_{n_2-1}}dt_{n_2}\;.
\ee

As we shall discuss in the next section, the expansion \eqref{eq:fjexp2} is a version of the 
Wild sums in the context of the hard sphere dynamics.

\bigskip
We expect $\langle K \rangle_t $ to be bounded for a fixed $t$ and exponentially growing in $t$, 
so that it is natural to introduce the quantities
\be
r_+=\limsup_{t \to \infty} \frac 1 t \log \langle K \rangle_t
\label{rate+}
\ee
and
\be
r_-=\liminf_{t \to \infty} \frac 1 t \log \langle K \rangle_t\;.
\label{rate-}
\ee

Note that $r_{\pm}$ are  computed by using the macroscopic scale of times, in which
the mean flight time is $O(1)$. We stress that the introduced quantities refer only to the kinetic 
reduced description, given by the homogeneous Boltzmann equation. 
The corresponding quantities at the level of the particle system ($\e > 0$) are very difficult to handle with.
In particular we have no results stating that such quantities are equivalent to \eqref{rate+}, \eqref{rate-}
in the Boltzmann--Grad limit. In the last section, we shall present related numerical 
simulations.

As we shall discuss later on, generally speaking we expect that $r_+=r_-=r$ defined as in \eqref{rate}.

The quantities we have introduced make sense also at equilibrium, 
namely when $f_0=M_\b$ is a uniform Maxwellian with inverse temperature $\b>0$.
Presently we are not able to show, even in this case, that $r_-=r_+=r$. If this is true, observe that 
$r = r_\b$ depends only on the temperature 
(or the energy) of the Maxwellian $M_\b$.  On the other hand, by virtue of the $H$ Theorem, 
any (non equilibrium) distribution $f_0$ with the same energy should have the same value
of $r$. In the last section we will show some numerical evidence of this behaviour for the hard sphere 
system.

We observe further that the dependence on the temperature should be given by
\be
r_\b = \frac{r_1}{\sqrt\b}\;.
\label{eq:rescrb}
\ee
This follows from $M_\b(v) = \b^{3/2}M_1(\sqrt\b v)$ which implies, by \eqref{nu} and \eqref{n},
$R^\b(v) = (1/\sqrt\b) R^1(\sqrt\b v)$ and $$f^{(n),\b}(v,t) = \b^{3/2} f^{(n),1}\left(\sqrt\b v,\frac{t}{\sqrt\b}\right)\;.$$
Here we have used an upper index to indicate the dependence on the temperature of the corresponding 
quantities. The last equation can be obtained easily from \eqref{n} by a rescaling of all the integration
variables (times and velocities). It follows that $\langle K \rangle^\b_t = \langle K \rangle^1_{t / \sqrt\b}$,
so that \eqref{eq:rescrb} holds if $r_\b$ exists.

\section{A simple model} \label{sec:Max}
\setcounter{equation}{0}    
\def\theequation{3.\arabic{equation}}

In this section we briefly analyze a simplified model of the Boltzmann equation for Maxwellian molecules
with angular cut--off \cite{Bob88}, for which the computations of the mean cluster size $\langle K \rangle_t$ 
can be made explicitly. 

We consider the Boltzmann equation
\be
\pa_t f= J(f,f)-f
\label{BEM}
\ee
where
\be
J(f,f)(v)=
\int_{\RRR^3\times S^2} dv_1 d\o \ g\left(\cos\theta\right) f(v_1')f(v')
\ee
for some nonnegative function $g$ satisfying $g=0$ for $\cos\theta < 0$, 
and $$\cos\theta = \frac{(v-v_1)}{|v-v_1|}\cdot\o\;.$$
Note that we have fixed the time scale in such a way that 
\be
\int_{ S^2}  d \o \,   g(\cos\theta)=1\;.
\label{norm}
\ee

Proceeding as in the previous section, we write the associated hierarchy
\be
\pa_t f_j = J_{j+1}f_{j+1}- j f_j\;,
\label{hie2}
\ee
where $J_{j+1}$ is defined as $\CC_{j+1}$ (see \eqref{eq:defBco}) with the function $(v_k-v_{j+1})\cdot\o$
replaced by $g\left(\frac{(v_k-v_{j+1})}{|v_k-v_{j+1}|}\cdot\o\right)$.
Again $f_j=f^{\otimes j}$ where $f=f(v,t)$ solves \eqref {BEM}. The initial condition for \eqref{hie2} is $f_0^{\otimes j}$.

From \eqref{hie2} one deduces (the analogous of \eqref{eq:fjexp2} for $j=1$)
\bea
&& f(v,t) = \sum_{n\geq 0} \sum_{\Gamma_n(1)} \int_0^t dt_1  \cdots \int_0^{t_{n-1}}dt_n \int_{\RRR^{3n}}
dv_{2} \cdots dv_{1+n} \int_{S^{2n}} d\o_1 \cdots d\o_n
\nn \\ && \ \ \ \ \ \ \ \ \ \ \ \ \ \ \ \ \  \cdot g(\cos\theta_1)\cdots g(\cos\theta_n)
e^{-(t-t_1)} e^{-2(t_1-t_2)}\cdots e^{-(n+1)t_n} f_{0}^{\otimes(1+n)} (\bv^{n})\nn\\ \label{exp2}
\eea
where $\cos\theta_i = \frac{(v_{k_i} - v_{1+i})}{|v_{k_i} - v_{1+i}|}\cdot\o_i$.
Note that this coincides with the Wild sums introduced in \cite{Wi51}, see also \cite{Mc66,CCG00}.

The integral of the $n-$th term in \eqref{exp2} is
\bea
&& \int dv\, f^{(n)}(v,t) \nn\\
&& = e^{-t} \sum_{\Gamma_n(1)} \int_0^t dt_1 \, e^{-t_1}  \cdots \int_0^{t_{n-1}}dt_n\, e^{-t_n} 
\int_{S^{2n}} d\o_1 \cdots d\o_n \int_{\RRR^{3(n+1)}} d\bv_{1+n} 
\nn \\ && \ \ \ \ \ \ \ \ \ \ \ \ \ \ \ \ \ \ \ \ \ \  \cdot g(\cos\theta_1)\cdots g(\cos\theta_n)
f_{0}^{\otimes(1+n)} (\bv^{n})\nn\\
&&= e^{-t} \sum_{\Gamma_n(1)} \int_0^t dt_1 \, e^{-t_1}  \cdots \int_0^{t_{n-1}}dt_n\, e^{-t_n} 
\int_{S^{2n}} d\o_1 \cdots d\o_n \int_{\RRR^{3(n+1)}} d\bv^{n} 
\nn \\ && \ \ \ \ \ \ \ \ \ \ \ \ \ \ \ \ \ \ \ \ \ \  \cdot g(\cos\theta_1)\cdots g(\cos\theta_n)
f_{0}^{\otimes(1+n)} (\bv^{n})\nn\\
\eea
where we applied repeatedly $dv_i'dv_k'=dv_idv_k$ in the collision between particles $i$ and $k$ for a fixed impact vector
$\o$. Using the normalization of $f_0$ and \eqref{norm}, and computing the time integrations, we easily arrive to
\be
\int dv f^{(n)} (v,t)=e^{-t} \Big ( \int_0 ^t e^{-s} ds \Big ) ^n\;.
\ee

Therefore we conclude that
\be
\langle K \rangle_t =\sum_{n\geq 0} n \int dv f^{(n)} (v,t)= e^t-1\;.
\label{meanM}
\ee
In particular, $$r=1\;.$$

\section{Estimate of the mean cluster size for hard spheres} \label{sec:estimate}
\setcounter{equation}{0}    
\def\theequation{4.\arabic{equation}}

We observe preliminarily that there is an important difference between the expansion 
\eqref {eq:fjexp2} for hard spheres and the corresponding expansion \eqref {exp2} for Maxwell molecules. 
The first is an equation in the unknown $f$. Indeed in the expression of $R$, the $f$ {\em itself} 
appears. Conversely, the Maxwellian expansion yields the explicit solution in terms of the initial datum $f_0$. 
In particular, the control of \eqref{meancl} cannot work simply by direct computation as in the 
previous section. Furthermore the proof that the series defining $\langle K \rangle_t$ for the hard sphere system is 
absolutely and uniformly convergent, works for a sufficiently small time only \cite{La75}.

In what follows we shall obtain information on $\langle K \rangle_t$
by computing the time derivative of $f^{(n)}$ given in \eqref {n}. In this way we manage to
exploit conservation laws, exact compensations and the known properties of the solution 
to the homogeneous Boltzmann equation. 

Let us take the derivative of $f^{(n)}(t)$ defined by \eqref{n}:
\bea
&& \pa_t f^{(n)} (t)=-R f^{(n)} (t)\nn\\
&&\ \ \ \ \ \ \ \ \ \ \ \ \ \ \ + \CC_2 \left(
\int_0^t dt_2  \cdots \int_0^{t_{n-1}}dt_n 
\, e^{-\int_{t_2} ^{t} ds R_{2}(s) } \CC_3
\cdots \CC_{1+n}e^{-\int_0^{t_n} ds R_{1+n}(s) } f_0^{\otimes(1+n)}\right)\nn\\
&&\ \ \ \ \ \ \ \ \ \ \  = -R f^{(n)} (t) + \CC_2 f^{(n-1)}_2 (t)\;,
\eea
having used \eqref{jn}. Applying \eqref{prod} and writing explicitly the collision operator,
one obtains the following differential hierarchy:
\be
\pa_t f^{(n)} (v,t)=-R f^{(n)} (v,t) +
\sum_{n_1=0}^{n-1}  \int_{\RRR^3\times S^2_+} dv_1 d\o \ 
(v-v_1)\cdot\o \Big\{f^{(n_1)} (v_1',t)f ^{(n-1-n_1)}(v',t)\Big\}\;.
\label{diff}
\ee

Setting
\be
{\cal K} (v,t)=\sum_{n=0}^{\infty}  n\,  f^{(n)} (v,t)\;,
\label{got}
\ee
it follows formally
\bea
\label{eqgot}
&& \ \ \  \ \  \  \ \ \pa_t {\cal K} (v,t)=  -R\,{\cal K} (v,t)+\\
&& \sum_{n_1=0}^{\infty}  \sum_{n_2=0}^{\infty} (n_1 +n_2 +1)
\int_{\RRR^3\times S^2_+} dv_1 d\o \ (v-v_1)\cdot\o \Big\{f^{(n_1)} (v_1',t)f ^{(n_2)}(v',t)\Big\} \nn \\
&& =  -R\,{\cal K} (v,t)+ \int_{\RRR^3\times S^2_+} dv_1 d\o \ (v-v_1)\cdot\o   \nn \\
&& \Big\{{\cal K} (v',t)f (v_1',t) +f(v',t)  {\cal K} (v_1',t) +f(v',t)f (v_1',t)
\Big\}\;. \nn 
\eea
Note that the above integral includes a positive collision operator linearized around $f$,
plus an inhomogeneous term given by a positive collision operator acting on $f^{\otimes 2}$.

Now we define
\be
K_0=\int dv \, {\cal K} (v, t)=\langle K \rangle_t \,, \qquad  K_2=\int dv \, {\cal K} (v, t)  \, v^2\;.
\label{mom}
\ee
Using \eqref {eqgot} and \eqref{nu},
\bea
\label{der}
\frac d{dt} K_2=&&-\p \int dv \int dv_1 v^2 |v-v_1| f(v_1) {\cal K} (v) \\
&&+\int dv \int dv_1 \int_{S^2_+} d\o\,  \o \cdot (v-v_1) v'^2 \Big ( {\cal K} (v) f(v_1) +{\cal K} (v_1) f(v) \Big) \nn \\
&&+ \int dv \int dv_1 \int_{S^2_+} d\o \,  \o \cdot (v-v_1) v'^2 f(v_1)f(v)  \nn. 
\eea
Moreover, symmetrizing and using the energy conservation,
\bea
\label{A}
A_2 :=&&\int dv \int dv_1 \int_{S^2_+}  d\o\,  \o \cdot (v-v_1) v'^2 f(v_1)f(v)\\
&& = \frac{\pi}{2} \int dv \int dv_1 |v-v_1| (v^2+v_1^2)  f(v_1) f(v) \nn \\
 && \leq \p \| f \|_3^2\;,  \nn
\eea
where
$$
\| f\|_s :=\int dv f(v) (1+v^2)^{\frac s 2}\;.
$$
Similarly, the second term in the right hand side of \eqref{der} can be written as
\bea
&& \int dv \int dv_1 \int_{S^2_+} d\o\,  \o \cdot (v-v_1) \, v'^2 \Big ( {\cal K} (v) f(v_1) +{\cal K} (v_1) f(v) \Big)\nn\\
&& =  \int dv \int dv_1 \int_{S^2_+}  d\o\, \o\cdot (v-v_1)\, ( v^2+v_1^2 ) \,{\cal K} (v) f(v_1)\nn\\
&& = \p\int dv \int dv_1 \, |v-v_1|\, ( v^2+v_1^2 ) \,{\cal K} (v) f(v_1)\;.
\eea
Notice that the first term above cancels exactly the first term in the r.h.s. of \eqref{der}.
In conclusion:
\be
\frac d{dt} K_2=\p\int dv \int dv_1 v_1^2 |v-v_1| f(v_1) {\cal K} (v)+A_2\;.
\ee

With a similar computation we obtain
\bea
\label{deriv0}
\frac d{dt} K_0&&=\p\int dv \int dv_1  |v-v_1| f(v_1) {\cal K} (v)+A_0\; \\ \nn
&& =\int dv \, R(v)\,  {\cal K} (v)+A_0,
\eea
where
\bea
\label{A0}
A_0 := &&  \p\int dv \int dv_1  |v-v_1| f(v_1) f(v)\nn \\
 &&=\int dv \, R(v)\,  f(v) \;.
 \eea

We observe now that, if the initial datum has finite norm $\| f_0 \|_3$, then
$\| f(t) \|_3$ remains bounded at any positive time. This is shown for instance in Theorem 1.1 of \cite{MW99}
(and proved already in \cite{Des93}). In the same assumptions, putting 
$C_1 = \pi \sup_{t\geq 0}\| f(t) \|_3$ and $C_2 = 2\pi \sup_{t\geq 0}\| f(t) \|_3^2$, we get
\be
\label{dis}
\frac d{dt} K_2\leq C_1(\sqrt {K_0K_2} +K_0)+C_2
\ee
and
\be
\label{dis}
\frac d{dt} K_0\leq C_1 (\sqrt {K_0K_2} +K_0)+C_2\;.
\ee

Indeed,
\bea
&&\frac d{dt} K_2 \leq \p\int dv \int dv_1 v_1^2\left( |v|+|v_1| \right)f(v_1) {\cal K} (v)+A_2\nn\\
&& = \p \left( \int dv |v| {\cal K} (v) \right) \left( \int dv\, v^2\, f(v) \right) + 
\p K_0  \left( \int dv |v|^3 f(v) \right) 
+ A_2\nn\\
&& \leq \p \left( \int dv |v| {\cal K} (v) \right) \| f \|_3+ 
\p K_0  \| f \|_3 + \p \| f \|_3^2\;.
\eea
By Cauchy--Schwarz inequality, $\int dv |v| {\cal K} (v) \leq \sqrt{\int dv |v|^2 {\cal K} (v)}
\sqrt{\int dv {\cal K} (v)} = \sqrt{K_2K_0}$, hence
\bea
&&\frac d{dt} K_2 \leq \p \| f(t) \|_3\left(\sqrt {K_0K_2} +K_0\right)
+ \pi \| f(t) \|_3^2
\eea
which implies (3.12). To obtain the estimate (3.13) we follow the same path, but $A_0 \leq 2\p\left(\int
|v| f \right) \leq 2\p\| f(t) \|_3 \leq 2\p\| f(t) \|_3^2$.

Finally, to obtain a lower bound, we use that, if the initial datum $f_0$ has finite mass, energy
and entropy, then $f(t)$ is bounded from below by a Maxwellian for any $t>0$
(see e.g. \cite{PW97}). In particular
\be
\label{low}
R(v,t)\geq \tilde C
\ee
for some $\tilde C>0$ (depending on $f_0$). Therefore from \eqref{deriv0}--\eqref {A0} we obtain
\be
\frac d{dt} K_0\geq \tilde C ( K_0 + 1)\;.
\ee

Summarizing, we established the following:
\begin{thm} \label{thm:CL}
Let $f(t)$ be the solution of \eqref{BE} with initial datum $f_0$ 
such that $\| f_0 \|_3 < + \infty$ and $\int dv \, f_0(v)\log f_0 (v) < \infty$.
Then there exist positive constants $\ol C_1, \ol C_2, \tilde C$ such that
\be
\left(e^{\tilde C t}-1\right) \leq \langle K \rangle_t \leq \ol C_2 \left(e^{\ol C_1 t}-1\right)
\ee
for any $t\geq 0$. In particular, $r_+ \leq \ol C_1$ and $r_- \geq  \tilde C$.
\end{thm}
Note that the constant $\ol C_1$ is proportional to $\sup_{t\geq 0}\| f(t) \|_3$.
(for instance using $\sqrt {K_0K_2} \leq (K_0+K_2)/\sqrt{2}$, one finds $\ol C_1 = C_1(2+\sqrt{2})$).


\newcommand{\braket}[1]{\langle #1 \rangle}

\section{Numerical simulation} \label{sec:Numsim}
\setcounter{equation}{0}    
\def\theequation{5.\arabic{equation}}

The average size of backward clusters of a real hard sphere system 
is difficult to investigate mathematically and the agreement of its
behaviour with the predictions of Theorem \ref{thm:CL} is not obvious
a priori. In this section we carry out the molecular dynamics 
simulation for hard spheres and compare it with the above results. 
It turns out that $\langle K \rangle_t$ grows indeed exponentially.
The present simulations have to be considered as preliminary. 
A more detailed analysis will be presented in a forthcoming paper.

Let us explain the setting of our simulation.
We consider $N$ particles of diameter $\e$ 
confined in a cube of side $L$. 
The position and velocity of the $i-$th particle at time $t$ 
are denoted here by $x_i(t)$, $v_i(t)$, $i\in I_N=\{ 1,2, \cdots , N\}$.
At initial time $t=0$, the particles are 
uniformly distributed in the cube in such a way 
that they do not overlap each other. 
The initial velocities are independently distributed
according to a function $f_0$, which will be specified later. 
We let the particles evolve freely until 
either following two events occur: 
(i)  two of them collide with each other or
(ii) one of them undergoes elastic collision with the wall of the cube. 
The velocity of particle(s) involved in 
the event is changed according to the collision law. 
The above procedure is iterated 
until a given time $t$ is achieved. 

The sequence of times
$0<t_1<\cdots<t_m<\cdots<t_{m_c}<t$, $(m=1,2,\cdots,m_c)$ is 
defined here as the instants at which the collision between two 
particles occurs. 
During the simulation, we retain 
the pair of particles [say, a pair $(p_m,q_m)$] which undergoes a collision at time $t_m$. 
Therefore, at the end of simulation, 
we have $\{t_m\}$ and $\{(p_m,q_m)\}$ for $m = 1,\cdots,m_c$. 
Based on these quantities, we can obtain the backward cluster $J_i$ of 
a particle with index $i$, according to the definition given in Section \ref{sec:intro}.
Note that $J_i$ does not include $i$ itself, i.e., 
if the $i-$th particle does not collide with any particle, then $J_i$ is empty. 
Let us denote by $K_i$ the cardinality of the backward cluster $J_i$. 
Then, we define by $g_N(K,t)$ the distribution of $K_i$ at time $t$: 
\begin{align}
g_N(K,t) = N^{-1} \#
\{
i\in I_N \,|\, K_i(t) = K 
\}, \quad\quad \[\sum_{K=0}^{N-1} g_N(K,t) =1 \]. 
\label{fN}
\end{align}
The average of the cardinality is thus defined as
\begin{align}
\braket{K}_t= \sum_{K=0}^{N-1} K g_N(K,t). 
\label{Kt}
\end{align}

It may be worth showing that the quantity $g_N(k,t)$ is actually expected to be close to the quantity
$ \int f^{(k)} dv$ which we have  studied at the level of the Boltzmann equation. Indeed for a typical 
configuration $\bz_N$ and a fixed $t$
$$
g_N(k,t)=\frac 1 N \sum_i \chi_{ \{K_i(t)=k\} } ( \bz_N) \approx \frac 1 N \sum_i \langle \chi_{\{K_i(t)=k\} } \rangle\;,
$$
by virtue of the law of large numbers ($N$ large). Here  $\chi\{...\}$ is an indicator function and 
$ \langle \cdot  \rangle $ is the expectation with respect to the (almost factorized) initial 
distribution of the initial datum $\bz_N$. 
Moreover the Boltzmann--Grad limit yields
$$
\frac 1 N \sum_i \langle \chi_{\{K_i(t)=k\} } \rangle=\langle \chi_{\{K_1(t)=k\} }\rangle\approx  
\int f^{(k)}(v,t) dv.
$$

In accordance with the analysis, 
we fix $N\epsilon^2 = \lambda^{-1} =1$. Moreover, 
$L=1$. 
The initial velocities $v_i(0)$ $(i\in I_N)$ are 
generated according to the distribution $f_0$, 
which is, in the present simulation, 
\begin{subequations}\label{initial}
\begin{align}
&\text{Case 1:}\quad f_0(v) = f_{\infty}(v) \equiv
\frac{1}{(2\pi/3 )^{3/2}}\exp\(-\frac{|v|^2}{2/3}\), \quad 
\( E =\frac{1}{2}\), \label{case1}\\
&\text{Case 2:}\quad f_0(v) = \frac{1}{8}
\prod_{p=1,2,3}\chi\left\{\left|v^{(p)}\right|<1\right\}, \quad 
\(E =\frac{1}{2}\), \\
&\text{Case 3:}\quad f_0(v) = \frac{1}{(8\pi/3 )^{3/2}}\exp\(-\frac{|v|^2}{8/3}\), \quad 
\(E =2\), \label{case3}
\end{align}
\end{subequations} 
where $E=\int_{\R^3}\frac{1}{2}|v|^2 f_0(v) dv$ is the energy 
(we let the mass of particles be unity) and
$v^{(p)}$ is the $p-$th component of $v$.
Cases 1 and 3 are equilibrium states with different energy, while Case 2 is a nonequilibrium state having 
same energy as Case 1. 
The velocity distribution of particles in Case 2 approaches the 
equilibrium $f_{\infty}$ as time goes on. 
In the actual simulation, due to noise, 
the energy $E$ is not exactly identical to the assigned one.

\begin{figure}[bt]
\centering
\includegraphics[width=1\textwidth]{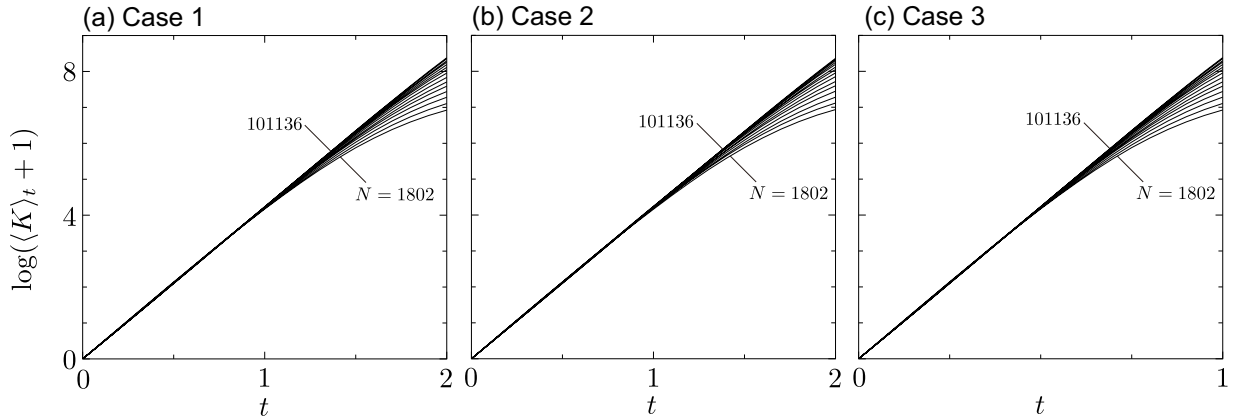}
\caption{
The average cardinality versus time 
in logarithmic scale 
for (a) Case 1, (b) Case 2, and (c) Case 3 [cf.~Eq.~\eqref{initial}]. 
Note that the range of $t$ in panel (c) is different from those in
panels (a) and (b).
For each curve, the ensemble average over $M$ different simulations is taken in order 
to decrease noise. 
We set $N\epsilon^2 = 1$ and $L=1$, while 
$(N,M)=(1802,72)$, 
$(2402,54)$, 
$(3203,40)$, 
$(4271,30)$, 
$(5695,23)$, 
$(7593,17)$, 
$(10125,13$, 
$(13500,10)$, 
$(18000,8)$, 
$(24000,6)$, 
$(32000,4)$,
$(42666,3)$, 
$(56888,3)$, 
$(75851,2)$, 
and 
$(101135,2)$. 
}
\label{fig1}
\end{figure}

Before stating the numerical results, it is necessary to mention 
the mean free time $\tau$ of the system. 
The mean free time $\tau$ at an equilibrium state (with energy $E$)
can be easily computed as 
$\tau = [4 (2\pi E/3 )^{1/2} N\epsilon^2]^{-1}$, see \cite{So07}.
Therefore, we obtain 
$\tau =(4\sqrt{\pi/3})^{-1}\approx 0.244$ for Cases 1 and 2, 
and $\tau = (8\sqrt{\pi/3})^{-1}\approx 0.122$ for Case 3. 
On the other hand, $\tau$ can be also computed from 
the numerical simulation. 
At the end of the simulation, we know $m_c$, 
which is the total number of collisions between particles. 
Since a single collision involves two 
particles, the total number of particles 
involved in $m_c$ collisions is $2 m_c$. 
The time--averaged free time is then 
$t/(2 m_c)$, during which 
one of the $N$ particles experiences a collision with one of the others. 
Thus, for a tagged particle, it takes 
$N t/(2 m_c)$ (on average) 
to experience a collision with one of the others. 
In the simulation, we have obtained 
$N t/(2 m_c)=0.242$ 
for Case 1, $N t/(2 m_c)=0.241$ for Case 2 
and $N t/(2 m_c)=0.121$ for Case 3 
when $N=101135$ and $t=2$.

\begin{table}[bt]
\begin{center}
{
\caption{The value of $\frac{1}{t}\log(\braket{K}_t+1)$}
\renewcommand{\arraystretch}{1.2}
{\setlength{\tabcolsep}{5pt}
   \begin{tabular}{c|ccc|ccc|ccc} \hline
   &
   \multicolumn{3}{c|}{Case 1} &
   \multicolumn{3}{c|}{Case 2} &
   \multicolumn{3}{c}{Case 3} \\ \hline
   $t$ $\backslash$ $N$
& 
   $1802$& $10125$& $101135$ &
   $1802$& $10125$& $101135$ &
   $1802$& $10125$& $101135$ 
\\ \hline
0.4 & 4.288 & 4.201 & 4.199 & 4.290 & 4.244 & 4.190 & 8.576 & 8.403 & 8.399\\
0.8 & 4.227 & 4.221 & 4.223 & 4.222 & 4.253 & 4.211 & 8.455 & 8.442 & 8.446\\
1.2 & 4.090 & 4.199 & 4.233 & 4.088 & 4.216 & 4.223 & 8.180 & 8.399 & 8.467\\
1.6 & 3.840 & 4.116 & 4.230 & 3.841 & 4.118 & 4.218 & 7.680 & 8.233 & 8.461\\
2.0 & 3.462 & 3.916 & 4.191 & 3.467 & 3.925 & 4.180 & 6.924 & 7.833 & 8.382\\
   \hline
   \end{tabular}}
\label{table1}
}
\end{center}
\end{table}

The plot in Fig.~\ref{fig1} and values in Table~\ref{table1} show that the
exponential behavior $\langle K \rangle_t \approx e^{rt} - 1$ is approached
as $N$ increases ($\e$ decreases), in a range of times including several
mean free flights. The value of $\frac{1}{t}\log(\braket{K}_t+1)$,
which should converge to $r$ as $N \to \infty$ and $t \to \infty$,
in Case 1 and that in Case 2 are almost coincident, 
as expected from the discussion before Eq. \eqref{eq:rescrb}. 
Observe that no transient is even visible in the case of a non--equilibrium, 
uniform distribution of velocities (Case 2). 
It is seen from Table~\ref{table1} that the value of $\frac{1}{t}\log(\braket{K}_t+1)$ in Case 3 
is almost twice larger than that of Case 1. 
This verifies \eqref{eq:rescrb} (here $\beta=3$ in Case 1 and 
$\beta = 3/4$ in Case 3).

\begin{figure}
\centering
\includegraphics[width=0.5\textwidth]{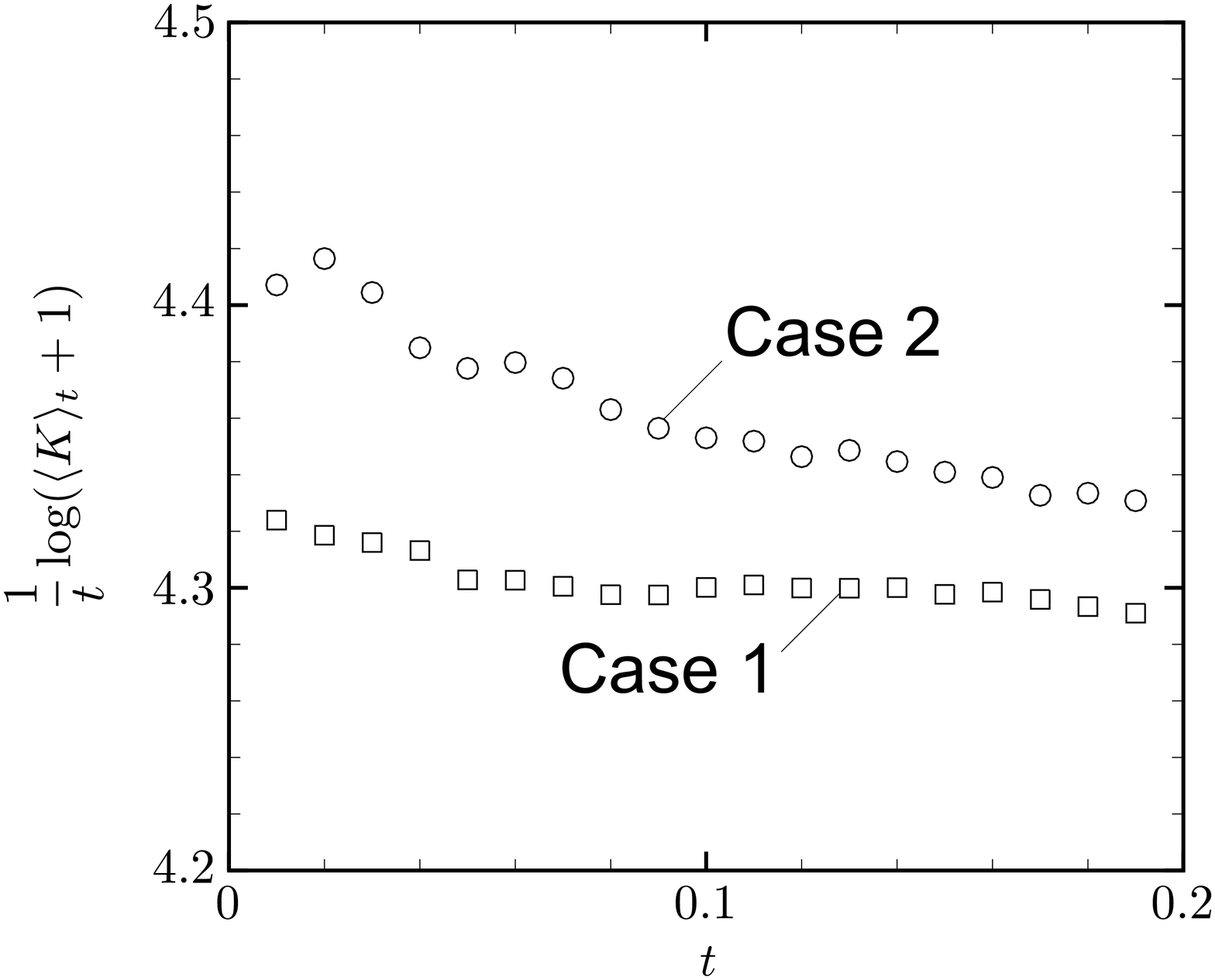}
\caption{The value of $\frac{1}{t}\log(\braket{K}_t+1)$ for 
Cases 1 (\textit{square}) and 2 (\textit{circle}) in the range of time 
less than one mean free time; $(N,M)=(1802,720)$. }
\label{fig02}
\end{figure}

Finally, we have checked whether $r$ is different between Cases 1 and 2, especially 
for small time $t\in[0,0.2]$. 
Note that $t\in[0,0.2]$ is within one mean free time at equilibrium $(\tau\approx 0.244)$. 
Figure~\ref{fig02} shows that the values of $\frac{1}{t}\log(\braket{K}_t+1)$ for both cases 
differ, but the discrepancy is small.

\newcommand{\red}[1]{\textcolor{red}{#1}}

\bigskip
\bigskip
\bigskip
\ni {\bf Acknowledgments.}
The authors would like to thank Chiara Saffirio and Herbert Spohn for stimulating discussions.
S. Simonella has been partially supported by PRIN 2009 ÒTeorie cinetiche e applicazioniÓ and 
by IndamÐCOFUND Marie Curie fellowship 2012, call 10.
T. Tsuji has been supported by the JSPS Institutional Program for Young Researcher
Overseas Visits.

\end{document}